\documentclass{svproc}
\usepackage{cite}
\usepackage{amssymb,amsfonts}
\usepackage{algorithm}
\usepackage{algpseudocode}
\usepackage{graphicx}
\usepackage{textcomp}
\usepackage{xcolor}
\usepackage{graphicx}
\usepackage{cite}
\usepackage{float}
\usepackage{physics}
\usepackage{amsmath}
\usepackage{qcircuit}
\usepackage{lscape}
\usepackage{indentfirst}
\usepackage{latexsym}
\usepackage{multirow}
\usepackage{tabls}
\usepackage{wrapfig}
\usepackage{longtable}
\usepackage{supertabular}
\usepackage{subfigure}
\usepackage{fancyvrb}
\usepackage{verbatimbox}
\usepackage{todonotes}

\usepackage{url}

\begin{document}
\mainmatter              
\title{A Quantum Algorithm To Locate Unknown Hashgrams}
\titlerunning{A Quantum Algorithm To Locate Unknown Hashgrams}  
%

\author{Nicholas R. Allgood\inst{1} \and Charles K. Nicholas\inst{1}}
\authorrunning{Nicholas R. Allgood et al.} 
%
\tocauthor{Nicholas R. Allgood, Charles K. Nicholas}
\institute{University of Maryland Baltimore County, Baltimore, MD 21250, USA
\email{allgood1@umbc.edu, nicholas@umbc.edu}}

\maketitle              

\begin{abstract}
Quantum computing has evolved quickly in recent years and is showing significant benefits in a variety of fields, especially in the realm of cybersecurity. The combination of software used to locate the most frequent hashes and $n$-grams that identify malicious software could greatly benefit from a quantum algorithm. By loading the table of hashes and $n$-grams into a quantum computer we can  speed up the process of mapping $n$-grams to their hashes. The first phase will be to use KiloGram to find the top-$k$ hashes and $n$-grams for a large malware corpus. From here, the resulting hash table is then loaded into a quantum simulator. A quantum search algorithm is then used search among every permutation of the entangled key and value pairs to find the desired hash value. This prevents one from having to re-compute hashes for a set of $n$-grams, which can take on average $O(MN)$ time, whereas the quantum algorithm could take $O(\sqrt{N})$ in the number of table lookups to find the desired hash values. 
\keywords{quantum computing, malware, n-gram, hashgrams, cybersecurity}
\end{abstract}
\section{Introduction}
Quantum computing is rapidly evolving and each day something new is being discovered. These discoveries are beginning making these concepts applicable across a variety of domains. In the late 1980's and early 1990's, quantum computing was entirely theoretical and many of the early algorithms created then have since provided a foundation on which to build other quantum algorithms.
While many of these algorithms, such as Simon's \cite{simon} and Grover's\cite{grover}, were seen as proof of concept algorithms, they in fact have more value on their own merits than simply providing a foundation for other algorithms. 

Though the situation is improving, one of the current limitations has to do with availability of quantum computing. While companies such as IBM\cite{ibmq} and D-Wave\cite{dwave} are providing access to their quantum computers at no cost via cloud platforms, they are still limited in the number of qubits and quantum volume available. For that reason, much of our work is done using Qrack\cite{qrack}, a high-performance quantum simulator. Simulators on classical hardware can simulate approximately 30 - 32 qubits.

One of the first steps in malware analysis is to perform \textit{static analysis} which searches the suspect binary file for static information such as strings that indicate the program's purpose, if a binary is maliciously packed, and whether the file is malicious.\cite{Sikorski}.  It is also desirable to compare the suspect binary with other binaries, malicious or not, to see if the suspect binary is similar to any of them.  An $n$-gram is a sequence of $n$ contiguous bytes, for some small integer $n$.  Files that happen to have many of the same $n$-grams, in roughly the same proportions, can be regarded as similar~\cite{Damashek1995}.  Historically, the value of $n$ might be in the range 2-6.  But unlike ordinary text, executable binaries use most if not all of the characters in the range {\tt 0x00} to {\tt 0xFF}.  For $n$ of 4, for example, that results in $256^4$ or roughly 4 billion possible $n$-grams to be tabulated.  More recently, as described below, larger values of $n$ are also of practical value, but in tabulating $n$-grams for any $n$ larger than say 3 or 4, a hash table would be used to keep track of which $n$-grams have been seen, and how often.  Hash tables are usually sized so that collisions don't matter too much in practice.  As a file is ingested, though, a lot of $n$-grams are seen multiple times, and the same hash value is computed multiple times. We will show how to improve $n$-gram tabulation by calculating an $n$-gram's hash once, storing the result, and using quantum search to find the desired hash value, without recomputing it, should that $n$-gram be seen again. 
This paper is organized as follows: in Section 2 we provide a review of related work.  We present the concept of quantum search as applied to $n$-grams in Section~\ref{sec:quantumSearch}.  Our numerical and simulation results are presented in Sections~\ref{sec:quantResults} and \ref{sec:simResults}.  In Section~\ref{sec:conclusion} we summarize our results and make suggestions for future work.

\section{Related Work}
\subsection{$n$-grams for Malware Analysis}

Cybersecurity professionals are constantly under pressure to identify and neutralize incoming threats. 
While antivirus software is essential, it is not always able to keep up with the threat.
Often a new piece of malware is released and performs some sort of damage before its signature is identified and updates made in the antivirus databases. 
Leveraging the latest techniques in machine learning, static analysis of malicious software has become a great tool in the arsenal against malware. \cite{Sikorski}
A large variety of malware is in the form of PE32 executable's that 
target the Microsoft Windows operating systems. 
One example use of $n$-grams would be to take sequences of bytes from a PE32 executable to construct features to be utilized by machine learning algorithms\cite{Shalaginov_2018}. Once such sequences of bytes are identified, the feature selection process goes through and eliminates duplicate or irrelevant pieces of information from the sets of data.  Using $n$-grams as features has proven effective in malware detection, showing up to a $97\%$ detection rate\cite{Shalaginov_2018}. 

There are a number of machine learning techniques used for malware detection
\cite{Shalaginov_2018}. Using $n$-grams as features are what makes it possible to leverage automated and intelligent classification methods. $n$-grams can be used for data, but also to represent a sequence of opcodes, as well as operating system API calls such as \textit{AdjustTokenPrivileges} for Win32 and \textit{execve} for Linux.

\subsection{KiloGram}

KiloGram\cite{Kilograms_2019} was 
released as open source software in 2020.\footnote{
\url{https://github.com/NeuromorphicComputationResearchProgram/KiloGrams}
} KiloGram takes a set of benign and known malicious software as input data.  The output will be a list of the top-$k$ most frequent $n$-grams found that are contained within the malicious software. Benign software is any software that is considered to not contain any malicious code where malware is any software that is design to cause harm in some fashion.
We chose the KiloGram approach since it can be used for a large number of $n$-grams and large values of $n$.
Of use to us, the KiloGram algorithm can handle $n$grams that are 8-bytes or larger while keeping 1000 or more of the most frequent entries. 

In the context of malware analysis, $n$-grams are used to represent strings that appear in some if not all members of a set of suspected malware specimens.  These $n$-grams then can be provided to other algorithms for a variety of uses, such as classification into malware families. KiloGram was designed with these uses in mind.
Recall that the $n$ in $n$-gram refers to some small integer $n$. For example, if we wish to process a 4-byte string such as {\tt 0xABCD}, you would see this called a $4$-gram. Unfortunately one major drawback of an $n$-gram based approach for malware detection, is that the shorter the $n$-gram, the more likely you will also find the byte sequence also benign software, making your rate for false positives increase. Fortunately, KiloGram was also designed to overcome this limitation by allowing the storing of larger and more specific $n$-grams, increasing the likelihood they will be unique within a variety or family of malware.

\subsection{Grover's Algorithm}

Grover's algorithm\cite{grover} was one of the first quantum searching algorithms to be developed.
Grover's has even been the inspiration for other quantum algorithms such as Shor's\cite{shor} factoring algorithm. While much attention and research has been specifically around Shor's algorithm with regards to quantum cryptography, Grover's has been used and even improved upon in recent years\cite{wang2017quantum}.

Grover's search algorithm implements what is known as an amplitude amplification algorithm\cite{Brassard_2002} which has been said to be a generalization of Grover's algorithm (although amplitude amplification was first discovered in 1997 by Gilles Brassard in 1997, and then a year later by Lov Grover). The fundamental idea is to increase (amplify) the probabilities of the desired results, and this is accomplished by using a sequence of reflections.\footnote{\url{https://docs.microsoft.com/en-us/quantum/libraries/standard/algorithms}} 
What is occurring in the amplitude amplification is that the reflections are rotated closer to the desired quantum state along the Bloch Sphere. The target state is marked as $\sin^2(\Theta)$ so that when the amplitude amplification algorithm is applied $m$ times, the probability of obtaining the correct state is $\sin^2((2m+1)\Theta)$ In other words, we think of the target state on the Bloch Sphere\cite{PhysRev.70.460} and we keep rotating it until we find the correct result, with each rotation getting slightly closer. 

\section{Quantum $N$-gram Searching}
\label{sec:quantumSearch}

\subsection{Amplitude Amplification}
Referring back to the previous statements, we explain that instead of looking up a value by key, we do a direct lookup by value. The reason being is we essentially have to invert the key/value lookup problem when dealing with quantum entanglement. 
Grover's search\cite{grover} makes heavy use of quantum entanglement. What this algorithm will do is when a lookup table is loaded into a quantum machine, Grover's algorithm will entangle all permutations of potential key and value pairs based upon the input. The next step is to perform what is known as amplitude amplification to the entangled pieces of data. Prior to the actual amplitude amplification, the oracle is queried which places a \textit{tag} value equal to our search value
As part of amplitude amplification, a \textit{tag} value that equals the search value is placed into memory and then the phase (sign) is flipped.

While amplitude amplification may sound like a phrase belonging in signal processing, it is heavily used in quantum mechanics to describe the nature of things, and it happens that most of those things happen to be analogue. For practical purposes, in quantum computing amplitude amplification and phase flipping refer to changing the sign of a value. For example, say we look at the following matrix and we wish to locate the value at row 1, column 3:

\[
\begin{bmatrix}
AB & CD & \textbf{EF} \\
12 & 97 & 85 \\
2D & 3F & 9C\\
\end{bmatrix}
\]

Once we perform the phase flip, we will get the following matrix:

\[
\begin{bmatrix}
AB & CD & \textbf{-EF} \\
12 & 97 & 85 \\
2D & 3F & 9C\\
\end{bmatrix}
\]

For our purposes, this \textit{tag} value is our $n$-gram we wish to locate and the key is the hash provided (which is also the index value). As mentioned, the key and value are entangled and with each lookup (iteration) of Grover's search, we can visualize the Bloch Sphere is rotated closer to the desired $n$-gram with each iteration.

Anything written about topics such as signal processing and quantum mechanics would be remiss if it failed to mention the Fourier Transform.\footnote{\url{https://www.encyclopediaofmath.org/index.php/Fourier_transform}}We are given a function $f(x)$ and the Fourier Transform breaks down $f(x)$ to its constituent frequencies\cite{quantum_fourier}. The conceptual structure of quantum mechanics defines the existence of pairs of complementary variables $p$ and $q$ connected by the Heisenberg uncertainty principle. We can measure a particle's quantum mechanical position, but by doing so we lose information about the particle's momentum\cite{quantum_fourier}. Going deeper into quantum mechanics, this gets into what is known as the wave-particle duality of nature, for which the physical state of a particle can be described by a wave function. The wave functions are used to describe the physical state of a particle and one can use either a function of $p$ or a function of $q$, but never both. The real vector space that is the set of all possible physical states and which contain the $p$-axis and $q$-axis is known as a \textit{phase space}.

Referring back to phase shifting and amplitude amplification as part of the algorithm, quantum mechanics choose a specific polarization of a defined space and picks a subspace containing half of its dimensions. In contrast to picking all of the points within this selected space that contains the q-axis, the quantum Fourier transform
takes the set of all complex-valued wave functions on the axis\cite{quantum_fourier}. We then examine the p-axis which while also having a valid polarisation, has a set of possible states of a particle related to the first representation by the Fourier transform:

\begin{equation}
    \Phi(p) = \int\psi(q)^{2\pi i \frac{pq}{h}}dq
\end{equation}

Physical states exist inside what is known as an $\mathcal{L}^2$ space, which is a vector space (specifically a measure space) that contains all of the squarable integral functions. Due to this property, an $\mathcal{L}^2$ space is more specifically a Hilbert space\cite{ruden_analysis}. According to Plancherl's theorem,\footnote{\url{https://link.springer.com/article/10.1007\%2FBF03014877}} Fourier Transforms also exist inside $\mathcal{L}^2$ spaces. Quantum mechanical operators are required to be unitary and a Fourier Transform within a $\mathcal{L}^2(R_n)$ space applied to itself is unitary. This upholds the unitary requirement for all quantum computing operations.

\section{Quantitative Results}
\label{sec:quantResults}

\subsection{Grover's Circuits}

Grover's algorithm is an \textit{oracle} based algorithm and in the majority of the literature that discusses Grover's algorithm, it's typically split into four parts:
\begin{enumerate}
    \item Initialization 
    \item Oracle processing
    \item Amplitude amplification
    \item Measurement
\end{enumerate}
We now describe
how a quantum simulator, in particular Qrack\cite{qrack}, implements both the oracle and amplification components of Grover's search.
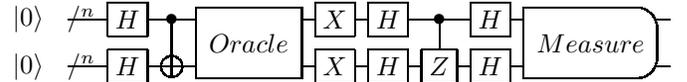
\begin{figure}[H]
    \centering 
    \Qcircuit  @C=1.1em @R=1.2em{
& & \ket{0} & & & & {/^n}\qw & \qw & \gate{H} & \ctrl{1} & \multigate{1}{Oracle} & \gate{X} & \gate{H} & \ctrl{1} & \gate{H} & \multimeasureD{1}{Measure} & \qw \\
& & \ket{0} & & & & {/^n} \qw & \qw & \gate{H} & \targ & \ghost{Oracle} & \gate{X} & \gate{H} & \gate{Z} & \gate{H} &  \ghost{Measure} & \qw  \\\\
& & Cl.\ Reg & & & & & {/^n}\cw & \cw & \cw & \cw & \cw & \cw & \cw & \cw & \cw & \cw
} 
    \caption{Example Grover's Circuit}
    \label{fig:grovers_circuit}
\end{figure}

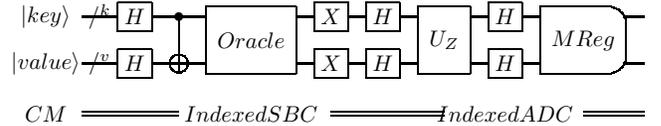
\begin{figure}[H]
    \centering
    \scalebox{0.8} {
    \Qcircuit @C=1em @R=1em{
      \lstick{\ket{key}} & {/^k} \qw & \gate{H} & \ctrl{1} & \multigate{1}{Oracle} & & \gate{X} \cwx[1] & \gate{H} & \multigate{1}{U_Z} & \gate{H} \cwx[1] & &  \multimeasureD{1}{MReg} & \qw \\
      \lstick{\ket{value}} & {/^v} \qw & \gate{H} & \targ & \ghost{Oracle} \cwx[1] & & \gate{X} \cwx[1] & \gate{H} & \ghost{U_Z} & \gate{H} \cwx[1] & & \ghost{MReg} \cwx[1] & \qw \\
     \lstick{Cl.\ Mem} & \cw & \cw & \cw & \gate{IndexedSBC} & \cw & \cw & \cw  & \cw & \cw & \gate{IndexedADC} & \cw & \cw & \cw 
    }
    }
    \caption{Qrack Implementation Grover's Circuit}
    \label{fig:grover_qrack}
\end{figure}


Figure \ref{fig:grovers_circuit} describes a quantum circuit for a standard Grover's search implementation. Figure \ref{fig:grover_qrack} shows Qrack's implementation of the quantum circuit for Grover's search over a key and value pair. A few comments on the notation: ${/^n}$ is shorthand to state that each of the gates apply to $n$ qubits. In the Qrack example, we chose ${/^k}$ and ${/^v}$ to represent the number of qubits used for the key and value. We use a $U_z$ to represent a phase-flip operation. We chose this over the standard Pauli-Z gate since we want a single permutation's phase flipped instead of flipping the phase on every individual $\ket{0}$. In the Qrack implementation, we are applying the phase-flip to all of the used qubits simultaneously. That is, we start out with setting our qubit permutations to $\ket{0}$. Next we apply a Hadamard gate to each of the qubits to place them into superposition where each qubit now equals $\frac{1}{\sqrt{2}}\ket{0} + \ket{1}$ and $\frac{1}{\sqrt{2}}\ket{0} - \ket{1}$. The second step of this circuit is to place all qubits through the oracle as defined in Grover's algorithm.
The next series of gates are CNOT gates which were previously place in superposition, the superimposed values of $\ket{1}$ will trigger a NOT operation on the target qubits. 
The IndexedSBC operation is in reference to Qrack's \textit{IndexedSBC}\cite{qrack_docs} operator that we will cover later in section \ref{sec:quantResults}.
We proceed with the phase-amplification part of the circuit by applying either an X or NOT gate on all qubits followed by both a Hadamard gate and a custom unitary phase-flip gate. Finally, we complete the circuit with another series of Hadamard gates, followed by IndexedADC which is Qrack's \textit{IndexedADC}\cite{qrack_docs} operation, and proceed with measurement of the resulting quantum state.

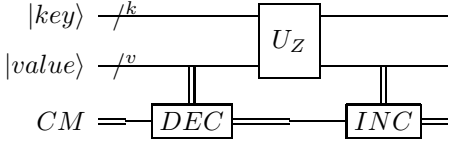
\begin{figure}[H]
    \centering
\[
\begin{array}{lc|r|}
\Qcircuit @C=3em @R=3em {
\lstick{\ket{key}} & \qw {/^k} & \qw \cwx[1] & \multigate{1}{U_Z} & \qw \cwx[1] \\
\lstick{\ket{value}} & \qw {/^v} & \qw \cwx[1] & \ghost{U_Z} & \qw \cwx[1] \\
\lstick{Cl.\ Mem} & \cw & \gate{DEC} & \cw & \gate{INC} & \cw
\\
} 
\end{array}
\]
    \caption{Qrack Implementation Grover's Oracle}
    \label{fig:grover_oracle}
\end{figure}

While the oracle might look small in comparison to the entire Grover's circuit, it's absolutely crucial to the algorithm. In the Qrack\cite{qrack} implementation of the oracle, we start by doing a \textit{DEC} operation for all qubits. This instruction is what starts the \textit{tag} process by subtracting the target value from a start value, 
typically $0$. For example, if our target value is $100$ then we would have $0-100 = -100$. From here we do $U_z$ gates on all qubits, flipping the phases of their respective amplitudes. Practically, this translates to flipping the sign of the bits so in the above example, this would make our value $+100$. Lastly, the oracle reverts the previous \textit{DEC} operation with Qrack's \textit{INC} operation to return to the original value, only with the sign flipped. To finalize our example, we add $0 + (+100)$ where $+$ is the phase, with our result being $+100$.

Theoretically, Grover's algorithm requires an average of $O(\sqrt{N})$ lookups to find a match for the specified target. While we are using a traditional lookup table for Grover's, the input time complexity evaluation might not be that obvious. If we dive into the bare fundamentals of Qrack/VM6502q, we notice we have a \textit{IndexedLDA} instruction\cite{qrack_docs}. This is a modified LOAD instruction that allows loading a key with a superimposed index into a quantum register. The \textit{IndexedLDA} operation is unitary by design so it will not affect the overall quantum state as it is loaded into the registers. The writing of the data with a superimposed index, will actually entangle the classical memory cache and the index register. Knowing this, we can say that the \textit{IndexedLDA} operation takes $O(1)$ to load data into quantum registers. In addition to the initial loads, there will be an input time complexity of $O\left(\sqrt{\frac{M}{N}}\right)$ where $M$ is the total number of keys in the lookup table and $N$ is the total number of matches\cite{nielsen00}. This yields an overall input time complexity of $O(1) + O\left(\sqrt{\frac{M}{N}}\right) = O\left(\sqrt{\frac{M}{N}}\right)$.

We use the term \textit{lookups} but this refers to the number of iterations of Grover's algorithm.  
To be specific, 
Qrack uses the following equation to determine the number of iterations to use\cite{qrack_docs}:

\begin{equation}
    floor\left[\frac{\pi}{4\arcsin^2(\frac{1}{\sqrt{2^N}})}\right]
\end{equation}

\section{Simulated Results}
\label{sec:simResults}

\subsection{Qrack Operations}
The Qrack\cite{qrack} implementation utilizes some specialized methods for implementing many of the 
operations in the oracle and amplitude amplification portions of the algorithm. 
Here are some of the most commonly used operations:\cite{qrack_docs}

\begin{description}
    \item \textbf{IndexedLDA:} Set 8 bit register bits by a superposed index-offset-based read from classical memory.
    \item \textbf{IndexedADC:} Add to entangled 8 bit register state with a superposed index-offset-based read from classical memory.
    \item \textbf{IndexedSBC:} Subtracts to entangled 8 bit register state with a superposed index-offset-based read from classical memory.
    \item \textbf{INC:} Integer addition without sign.
    \item \textbf{DEC:} Integer subtraction without sign.
    \item \textbf{H:} Hadamard gate implementation. 
    \item \textbf{ZeroPhaseFlip:} Controlled Z-gate implementation.
    \item \textbf{Z:} Z-gate implementation, non-controlled.
    \item \textbf{X:} X (NOT) gate implementation.
    \item \textbf{MReg:} Measures the current state of a quantum register(s).
\end{description}

\subsection{Benign vs. Malicious Datasets}

As briefly mentioned in Section~2, there are some limitations with quantum simulations, the most obvious being limited computing resources available for simulation. While Qrack\cite{qrack} can take full advantage of a GPU for processing using OpenCL\footnote{\url{https://www.khronos.org/opencl/}}, one typically is limited to simulating approximately 30-qubits. Qrack has some development branches of code where they are simulating 128-qubits for testing the quantum supremacy problem released by Google\cite{google_supremacy}, however, these branches are quite experimental.  To better appreciate why 30 qubits is a limitation for simulation, we must recall our base formula $2^n$ where $n$ is the number of qubits we wish to simulate. $2^n$ specifically refers to the total amount of quantum states we wish to simulate. With 30 qubits, we end up with $2^{30} = 1073741824$ or roughly one billion values.  But the amplitudes represented by the quantum states are complex numbers, so we must include the real and imaginary parts when factoring in memory requirements. We use $2^2$ bytes for the real value and $2^2$ bytes for the imaginary value. This then gives us $2^{2+2} = 16$  bytes for each of those one billion values, or 
\begin{equation}
    2^{30+4} = 17179869184\ bytes \approx 16GB
\end{equation}
\begin{table}[H]
    \centering
    \begin{tabular}{|l|l|l|l|}
    \hline
    Qubits & Real Bytes & Imaginary Bytes & Total Memory  \\
    \hline
    4 & 2 & 2 & 256 bytes \\
    8 & 2 & 2 & 4 KB \\
    16 & 2 & 2 & 1 MB \\
    24 & 2 & 2 & $\approx268$MB \\
    28 & 2 & 2 & $\approx4$GB \\
    30 & 2 & 2 & $\approx16$GB \\
    32 & 2 & 2 & $\approx64$GB \\
    40 & 2 & 2 & $\approx17$TB \\
    \hline
    \end{tabular}
    \caption{Simulation Memory Allocation}
    \label{tab:simulation_memory}
\end{table}

For our simulation, we used the following datasets created and used in the KiloGram project\cite{Kilograms_2019}:

\begin{table}[H]
    \centering
    \scalebox{.9}{
    \begin{tabular}{|l|l|l|l|}
    \hline
    Benign & Benign Files & Malicious & Malicious Files  \\
    \hline
    Windows 7 System32 & 4565 & Vxheaven 2015 & 284151
    \\
    MAML & 691 & VirusShare 2018 & 131072
    \\
    \hline
    \end{tabular} }
    \caption{Benign vs. Malicious Software Dataset}
    \label{tab:benign_malware_files}
\end{table}

\begin{table}[H]
    \centering
    \scalebox{.9}{
    \begin{tabular}{|l|l|l|l|}
    \hline
    Benign & Malicious & $n$-gram size & Kept $n$-grams \\
    \hline
    Windows 7 System32 & Vxheaven 2015 & 3 bytes & 64
    \\
    Windows 7 System32 & Vxheaven 2015 & 2 bytes & 16384
    \\
    Windows 7 System32 & Vxheaven 2015 & 2 bytes & 4096
    \\
    MAML & VirusShare 2018 & 3 bytes & 64
    \\
    MAML & VirusShare 2018 & 2 bytes & 2048
    \\
    MAML & VirusShare 2018 & 2 bytes & 1024
    \\
    \hline
    \end{tabular} }
    \caption{Benign vs. Malicious Software $n$-grams}
    \label{tab:benign_malware_ngrams}
\end{table}

The hardware and software used was a 16-Core Intel Xeon E5-2630 @ 2.4Ghz with 32GB RAM and two GeForce GTX 1660 video cards. The machine was running 64-bit Ubuntu Linux 18.04 and OpenCL 1.2. As one can see, due to the limitations of approximately 30-qubits, we had to select the number of bits for our key and value size with care. Since $n$-grams are typically byte sequences, we were limited to a maximum of $n$-grams with $n$ set to 3. Using $3$-grams gave us 24-qbits for our $n$-gram value with 6-qbits remaining for our index values. Utilizing a $2$-grams gave us a much larger span of bits to use for our index value (14-qbits). Recall that the index for this is the hash for a specific $n$-gram, and since KiloGram utilizes Rabin-Karp hashing \textit{modulo} $B$ where $B$ is the KiloGram bucket size\cite{Kilograms_2019}.

\begin{table}[H]
    \centering
    \begin{tabular}{|l|l|}
    \hline
    Number of $n$-grams & Number of Lookups (iterations)  \\
    \hline
    64 & $\sqrt{64} = 8$
    \\
    128 & $\sqrt{128} = 11.31 \approx 12$
    \\
    256 & $\sqrt{256} = 12$
    \\
    512 & $\sqrt{512} = 22.63 \approx 23$
    \\
    1024 & $\sqrt{1024} = 32$
    \\
    2048 & $\sqrt{2048} = 45.25 \approx 46$
    \\
    4096 & $\sqrt{4096} = 64$
    \\
    8192 & $\sqrt{8192} = 90.50 \approx 91$
    \\
    16384 & $\sqrt{16384} = 128$
    \\
    \hline
    \end{tabular}
    \caption{Grover's Lookups For $n$-gram Sizes}
    \label{tab:grovers_lookup_ngram}
\end{table}

As we can see from Table~\ref{tab:grovers_lookup_ngram}, more $n$-grams requires more iterations and the number of iterations increases by a much smaller amount as we keep a larger number of $n$-grams.
Using a practical example, below we describe pseudo-code for the Qrack implementation of Grover's algorithm in addition to showing the output for a 2-byte $n$-gram with a 10-bit index. We search for a $n$-gram with the value of \texttt{0xF3D7} which has an unknown hash, which we quickly find to be \texttt{0x3a9}.

\subsection{Example Hash Retrieval for $N$-gram: {\tt 0xf3d7}}

Table~\ref{tab:chances} is an example where we search for an $n$-gram with the value of \texttt{0xF3d7} that has a hash value of \texttt{0x3a9}.

\begin{figure}[H]
\begin{verbatim}
         0> chance of match:0.00876619
         1> chance of match:0.0242241
         2> chance of match:0.0471087
        ...
        22> chance of match:0.98967
        23> chance of match:0.998456
        24> chance of match:0.999461
After measurement (of value, key, or both):
Chance of match:1
Ngram: f3d7
Hash: 3a9
Total Iterations: 25
\end{verbatim} 
\caption{Searching for the hash of $n$-gram {\tt 0xF3D7}}
\label{tab:chances}
\end{figure}

\subsection{Qrack Pseudo-code}

In the following algorithm, we show pseudo-code utilizing Qrack that is an implementation of both an oracle and amplitude amplification for Grover's search.

\begin{algorithm}[H]
  \begin{algorithmic}
        \State $idxLen =  10$
        \State $valLen =  16$
        \State $cryIdx = idxLen + valLen$
        \State $ngrams =  ngramtable[indexLength]$
        \State $ngram =  0xf3d7$
        \State $qReg =  CreateQuantumInterface(*params)$
        \State $qReg =  SetPermutation(0)$
        \State $qReg =  H(valLen, idxLen)$
        \State $qReg =  IndexedLDA(valLen, idxLen, 0, valLen, ngrams)$
    \Procedure{QueryOracle}{$tPerms,qReg,valueSt,valLen$}
            \State $qReg =  DEC(tPerms, valueSt, valLen)$
            \State $qReg =  ZeroPhaseFlip(tPrems, valueSt, valLen)$
            \State $qReg =  INC(tPrems, valueSt, valLen)$
    \EndProcedure 
    \Procedure{AmplitudeAmplifiation}{}
        \State $idxLen =  10$
        \State $valLen =  16$
        \State $cryIdx = idxLen + valLen$
        \State $ngrams =  ngramtable[idxLength]$
        \State $ngram =  0xf3d7$
        \State $qReg =  CreateQuantumInterface(params)$
        \State $qReg =  SetPermutation(0)$
        \State $qReg =  H(valLen, idxLen)$
        \State $qReg =  IndexedLDA(valLen, idxLen, 0, valLen, ngrams)$
        \For{$i  =  0$ to $ floor\left[\frac{\pi}{4\arcsin^2(\frac{1}{\sqrt{2^N}})}\right]$}\\ 
                \State $TagValue(ngram, qReg, 0, valLen)$
                \State $qReg =  X(cryIdx)$
                \State $qReg =  IndexedSBC(valLen, idxLen, 0, valLen,$
                \State $cryIdx, ngrams)$
                \State $qReg =  X(cryIdx)$
                \State $qReg =  H(valLen, idxLen)$
                \State $qReg =  ZeroPhaseFlip(valLen, idxLen)$
                \State $qReg =  H(valLen, idxLen)$
                \State $qReg =  IndexedADC(valLen, idxLen, 0, valLen,$
                \State $cryIdx, ngrams)$
        \EndFor \\ 
    \EndProcedure
  \end{algorithmic} 
\end{algorithm}

\section{Conclusion}
\label{sec:conclusion}

We have shown that combining the results of an efficient $n$-gram collection software such as KiloGram with quantum computing, we can provide a faster way of finding a previously computed, but currently unknown hash for a known $n$-gram. 
We have compared this solution to 
the classical approach, and have shown that for a large number of $n$-grams, the quantum based solution outperforms them substantially. When better quantum hardware is available, these concepts could be applied to cryptographic hashes such as SHA-256 or BLAKE3. 
We hope that our work will remain useful
when better quantum computers are available. Quantum computing research is continuing to grow each day and while it might seem that adequate enough hardware is far into the future, it is will be upon us before we realize and cybersecurity professionals will need to be ready.

\section*{Acknowledgment}
We extend our thanks to our colleagues Sam Lomonaco and Edward Raff for their comments on an earlier version of this paper.\cite{Allgood2020}. We also extend our sincere gratitude to Dan Strano for the development and support of the Qrack\cite{qrack} quantum simulator.
%
%
\nocite{*}
\bibliographystyle{unsrt} 
\bibliography{references}

\begin{thebibliography}{10}

\bibitem{simon}
D.R. Simon.
\newblock On the power of quantum computing.
\newblock In {\em Foundations of Computer Science, 1994 Proceedings., 35th
  Annual Symposium on: 116–123}, 1994.

\bibitem{grover}
Lov~K. Grover.
\newblock A fast quantum mechanical algorithm for database search.
\newblock {\em Proceedings of the twenty-eighth annual ACM symposium on Theory
  of computing - STOC ’96}, 1996.

\bibitem{ibmq}
IBM.
\newblock {IBM} quantum experience.
\newblock \url{https://quantum-computing.ibm.com}, 2020.

\bibitem{dwave}
D-Wave.
\newblock {D-wave}.
\newblock \url{https://dwavesys.com}, 2020.

\bibitem{qrack}
Daniel Strano and Benn Bollay.
\newblock {Qrack} a comprehensive, gpu accelerated framework for developing
  universal virtual quantum processors.
\newblock \url{https://github.com/vm6502q/qrack}, 2020.

\bibitem{Sikorski}
Michael Sikorski and Andrew Honig.
\newblock {\em {Practical Malware Analysis}}.
\newblock no starch press, 2012.

\bibitem{Damashek1995}
Marc Damashek.
\newblock {Gauging Similarity with N-Grams}.
\newblock {\em Science}, 267(5199):843--848, 1995.

\bibitem{Shalaginov_2018}
Andrii Shalaginov, Sergii Banin, Ali Dehghantanha, and Katrin Franke.
\newblock Machine learning aided static malware analysis: A survey and
  tutorial.
\newblock {\em Cyber Threat Intelligence}, page 7–45, 2018.

\bibitem{Kilograms_2019}
Edward Raff, William Fleming, Richard Zak, Hyrum Anderson, Bill Finlayson,
  Charles~K. Nicholas, and Mark Mclean.
\newblock {KiloGrams: Very Large N-Grams for Malware Classification}.
\newblock In {\em Proceedings of KDD 2019 Workshop on Learning and Mining for
  Cybersecurity (LEMINCS'19)}, 2019.

\bibitem{shor}
P.W. Shor.
\newblock Algorithms for quantum computation: discrete logarithms and
  factoring.
\newblock {\em Proceedings 35th Annual Symposium on Foundations of Computer
  Science, Santa Fe, NM}, pages 124--134, 1994.

\bibitem{wang2017quantum}
Yan Wang.
\newblock A quantum walk enhanced grover search algorithm for global
  optimization, 2017.

\bibitem{Brassard_2002}
Gilles Brassard, Peter Høyer, Michele Mosca, and Alain Tapp.
\newblock Quantum amplitude amplification and estimation.
\newblock {\em Quantum Computation and Information}, page 53–74, 2002.

\bibitem{PhysRev.70.460}
F.~Bloch.
\newblock Nuclear induction.
\newblock {\em Phys. Rev.}, 70:460--474, Oct 1946.

\bibitem{quantum_fourier}
D.~Coppersmith.
\newblock An approximate fourier transform useful in quantum factoring, 2002.

\bibitem{ruden_analysis}
Walter Rudin.
\newblock {\em Real and Complex Analysis, 3rd Ed.}
\newblock McGraw-Hill, Inc., USA, 1987.

\bibitem{qrack_docs}
Daniel Strano and Benn Bollay.
\newblock Vm6502q and qrack.
\newblock \url{https://vm6502q.readthedocs.io/en/latest/index.html}, 2020.

\bibitem{nielsen00}
Michael~A. Nielsen and Isaac~L. Chuang.
\newblock {\em Quantum Computation and Quantum Information}.
\newblock Cambridge University Press, 2000.

\bibitem{google_supremacy}
Frank Arute, Kunal Arya, and Ryan~Babbush et~al.
\newblock Quantum supremacy using a programmable superconducting processor.
\newblock {\em Nature}, 574:505–510, 2019.

\bibitem{rabin}
Thomas~H. Cormen, Charles~E. Leiserson, Ronald~L. Rivest, and Clifford Stein.
\newblock {\em Introduction to Algorithms}, pages 911--916.
\newblock MIT Press, second edition, 1990.

\bibitem{deutsch_jozsa}
David {Deutsch} and Richard {Jozsa}.
\newblock {Rapid Solution of Problems by Quantum Computation}.
\newblock {\em Proceedings of the Royal Society of London Series A},
  439(1907):553--558, December 1992.

\bibitem{Pang_2012}
Chao-Yang Pang, Ri-Gui Zhou, Cong-Bao Ding, and Ben-Qiong Hu.
\newblock Quantum search algorithm for set operation.
\newblock {\em Quantum Information Processing}, 12(1):481–492, Mar 2012.

\bibitem{coles2018quantum}
Patrick~J. Coles, Stephan Eidenbenz, and Scott~Pakin et~al.
\newblock Quantum algorithm implementations for beginners, 2018.

\bibitem{hashgrams}
Edward Raff and Charles Nicholas.
\newblock Hash-grams: Faster n-gram features for classification and malware
  detection.
\newblock In {\em Proceedings of the ACM Symposium on Document Engineering
  2018}, DocEng ’18, New York, NY, USA, 2018. Association for Computing
  Machinery.

\bibitem{entanglement}
Nicolas~J. Cerf and Chris Adami.
\newblock Information theory of quantum entanglement and measurement.
\newblock {\em Physica D: Nonlinear Phenomena}, 120(1-2):62–81, Sep 1998.

\bibitem{Brassard}
G.~Brassard and P.~Hoyer.
\newblock An exact quantum polynomial-time algorithm for simon’s problem.
\newblock {\em Proceedings of the Fifth Israeli Symposium on Theory of
  Computing and Systems}, 1997.

\bibitem{10.1007/978-3-642-22786-8_6}
Sachin Jain and Yogesh~Kumar Meena.
\newblock Byte level n--gram analysis for malware detection.
\newblock In K.~R. Venugopal and L.~M. Patnaik, editors, {\em Computer Networks
  and Intelligent Computing}, pages 51--59, Berlin, Heidelberg, 2011. Springer
  Berlin Heidelberg.

\bibitem{pednault2017breaking}
Edwin Pednault, John~A. Gunnels, Giacomo Nannicini, Lior Horesh, Thomas
  Magerlein, Edgar Solomonik, Erik~W. Draeger, Eric~T. Holland, and Robert
  Wisnieff.
\newblock Breaking the 49-qubit barrier in the simulation of quantum circuits,
  2017.

\bibitem{Allgood2020}
Nicholas~R. Allgood.
\newblock A quantum algorithm to locate unknown hashes for known n-grams within
  a large malware corpus, May 2020.

\end{thebibliography}
\end{document}